# Stabilizing Multimode Hopping Oscillations and Reducing Associated Noise in Long-Wavelength Laser Diode Using External Optical Feedback

Moustafa Ahmed[1,*], Ahmed Bakry[1], Ahmed Alshahrie[1] and Hamed Dalir[2]

[1]Department of Physics, Faculty of Science, King Abdulaziz University, 80203 Jeddah21589, Saudi Arabia

[2]Department of Electrical and Computer Engineering, George Washington University, Washington, DC 20052, USA

[*]Corresponding author: mostafa.farghal@mu.edu.eg

**Abstract**

We report on converting the multimode hopping oscillation (MHO) in long-wavelength semiconductor laser into single-mode oscillation (SMO) by applying external optical feedback (OFB). We characterize and compare the noise performance of the laser when supporting SMO and multimode oscillations. The study is based on a modified time-delay multimode rate-equation model of the laser that includes mechanisms of spectral gain suppression along with OFB induced due to multiple reflections by an external reflector. The study is applied to 1.55μm-InGaAsP laser that exhibits multimode hopping in its solitary version and supports wide bandwidth. The noise is evaluated in terms of the relative intensity noise (RIN). We show that when OFB synchronizes with the asymmetric gain suppression (AGS), it enhances the gain of one longer wavelength mode and supports SMO. In this case OFB improves the noise performance of the laser. On the other hand, when OFB works against AGS, it sustains hopping multimode oscillation (HMMO) and deteriorates the side-mode suppression ratio (SMSR) and the noise performance.

## 1. Introduction

Low cost, wide bandwidth, SMO and low noise are basic requirements on semiconductor laser for application in modern technological systems, for instance fiber communications. Because of the simple structure of semiconductor lasers with Fabry-Perot (FP) resonators, they are attractive for low-cost applications [1]. However, these lasers when emitting in the C-band of telecommunication (long wavelengths of 1.2 ~ 1.6μm) oscillate in multi-longitudinal modes because of the broad gain spectrum and the AGS that makes the gain spectrum shallow over a number of modes on the long-wavelength side [2-5]. These lasers happen to have anti-correlations and multimode hopping among such long-wavelength modes [6]. Therefore, they are proved in theory and seen in experiment to reveal an asymmetric-like multimode spectrum [5-7]. HMO was proved to deteriorate the noise performance of the laser; it enhances the noise level in the low-frequency regime compared to that of a single-mode laser [5,8], especially around the mode-hopping frequency.

Recently the authors reported that application of external OFB from a very close reflector could release the anti-correlations among the hopping modes, stabilize the laser operation and convert the multimode oscillations into oscillations in a single mode [9]. In this limit of the very short external cavity, the chaos dynamics cycles may disappear, and the laser operation becomes more stable [10]. The authors showed that this case of very-short external cavity is also advantageous to enhance the modulation bandwidth and speed of the laser due to the induced photon-photon-resonance effect between the modulating field and the external-cavity oscillating modes [11].

The noise properties of semiconductor lasers under OFB depends on the OFB parameters such as the feedback strength and length of the external cavity and the phase relationship between the internal field and re-injected field at the front facet [12]. Imran and Yamada [13] classified the multimode noise spectrum into mode-competition noise, low-frequency noise and flat (white) noise in the regime of low-noise frequencies. It is necessary for applications of the FP laser to specify both regimes of OFB that yield SMO with low and flat (white) noise levels and HMMO with enhanced noise levels.

In this paper, we present modeling on the noise properties of long-wavelength (1.55μm-InGaAsP) multimode laser subject to OFB from a very short external cavity. We are aimed at exploring the feedback regimes that correspond to SMO with low noise levels and HMMO with enhanced noise. The laser noise is evaluated in terms of the frequency spectrum of RIN. The model is based on a modified time-delay multimode stochastic rate-equation model that includes the various mechanisms of spectral gain suppression along with OFB due to multiple reflections by an external reflector. We clarify the role of AGS along with the Langevin noise source to induce multimode hopping in this type of multimode laser. We also investigate influence of OFB on the HMMO of this type of laser and consequently on the RIN spectrum. We show that when OFB synchronizes with AGS, it enhances the gain of one longer wavelength mode and supports SMO with low RIN.

On the other hand, when OFB releases AGS, it sustains the multimode hopping and deteriorates both SMSR and noise performance.

The structure of the paper is as follows. In section 2, we introduce the theoretical stochastic rate equation model under OFB used in simulation. The numerical procedures are given in section 3, while the simulation results on the modal oscillations and noise properties under OFB are presented in section 4. Finally conclusions appear in section 5.

## 2. Theoretical Model

In the multimode model of semiconductor lasers, the gain spectrum is described by

$$G_p = A_p - BS_p - \sum_{q \neq p} [D_{p(q)} + H_{p(q)}] S_q \tag{1}$$

where $A_p$ is the linear gain of mode $p$, which is described by the following parabolic function of wavelength $\lambda_p$ [5,8],

$$A_p = \frac{a\xi}{V}\left[N - N_g - bV(\lambda_p - \lambda_0)^2\right] \quad p = 0, \pm 1, \pm 2, ... \tag{2}$$

$a$ is the differential gain coefficient, $N_g$ is injected carrier number at transparency, and the $b$ is the width of the spectrum. $\lambda_0$ is the wavelength of mode $p = 0$, which is assumed to coincide with the peak wavelength of the gain spectrum. The wavelength of the other modes is $\lambda_p = \lambda_0 + p\Delta\lambda$ where $\Delta\lambda$ is the mode wavelength separation. In the above equation, the other terms of equation (1) represent the nonlinear gain coefficients contributing to gain suppression. The coefficient $B$ is the self-modal suppression coefficient, whereas $D_{p(q)}$ and $H_{p(q)}$ are coefficients of symmetric and asymmetric suppressions of mode $p$ by other modes $q \neq p$. $D_{p(q)}$ is described by the following Lorentzian dispersion relation around mode $p$ [5,8],

$$D_{p(q)} = \frac{4}{3} \frac{B}{\left(2\pi c \tau_{in}/\lambda_p^2\right)(\lambda_p - \lambda_q)^2 + 1} \tag{3}$$

where $c$ is the speed of light in free space and $\tau_{in}$ is the intraband relaxation time. $H_{p(q)}$ is called AGS and because it is asymmetric around mode $p$ [5,8,14],

$$H_{p(q)} = \frac{3}{8\pi}\left(\frac{a\xi}{V}\right)^2 (N - \overline{N}) \frac{\alpha \lambda_p^2}{\lambda_q - \lambda_p} \tag{4}$$

where $\alpha$ is the linewidth enhancement factor and $\overline{N}$ is the time-averaged value of $N$. For modes $q$ with $\lambda_q > \lambda_p$, $H_{p(q)}$ is positive and works to enhance these modes with longer wavelengths, whereas $H_{p(q)}$ suppresses the gain of modes with longer wavelengths. The amount of asymmetry is rather large for larger values of the $\alpha$-factor, such as the case of long-wavelengths (InGaAsP) lasers which is then characterized by strong mode coupling and competition [5,6,8,9,14].

OFB is treated as time delay of mode radiation at the laser front facet due to multiple round trips in the external cavity formed between the front facet of reflectivity $R_f$ and the external reflector of reflectivity $R_{ex}$. The period of each round trip is $\tau = 2n_{ex}L_{ex}/c$, where $n_{ex}$ and $L_{ex}$ are the refractive index and length of the external cavity, respectively. In this case, the laser rate equations of the injected carrier number $N$ as well as the photon number $S_p$ and phase $\theta_p$ of mode $p$ are modified to the following forms

$$\frac{dN}{dt} = \frac{I}{e} - \sum_p A_p S_p - \frac{N}{\tau_s} + F_N(t) \tag{5}$$

$$\frac{dS_p}{dt} = \left( G_p - \frac{1}{\tau_p} + \frac{c}{n_D L_D} \ln|U_p(t-\tau)| \right) S_p + C_p \frac{N}{\tau_s} + F_{Sp}(t) \tag{6}$$

$$\frac{d\theta_p}{dt} = \frac{1}{2}\left( \frac{\varepsilon a \xi}{2V}(N-\overline{N}) - \frac{c}{n_D L_D} \varphi_b \right) + + F_{\theta p}(t) \tag{7}$$

where the time-delay function $U_p(t-\tau)$ is defined as [9]

$$U_p(t-\tau) = |U_p|e^{-j\varphi_p} = 1 - \frac{1-R_f}{R_f} \sum_{m=1} \sqrt{R_f \eta R_{ex}}^m e^{-jm\omega\tau} \frac{S_p(t-m\tau)}{S_p(t)} \frac{e^{j\theta_p(t-m\tau)}}{e^{j\theta_p(t)}} \tag{8}$$

with

$$\varphi = -\tan^{-1}\frac{\mathrm{Im}\,U_p}{\mathrm{Re}\,U_p} + n\pi \quad (-\pi \leq \pi \leq \pi) \tag{9}$$

In equation (6), $\tau_p$ is the photon lifetime, and $C_p$ is the rate of inclusion of spontaneous emission into simulated emission of mode $p$, and is given in terms of the FWHM of the spectrum $\delta\lambda$ as [15]

$$C_p = \frac{1}{4\left(\frac{\lambda_p - \lambda_0}{\delta\lambda}\right)^2 + 1} \tag{10}$$

This model is generalization of the rate equation model in [13] to account multiple round trips in the external cavity and to include fluctuations of the carrier number in equation (5). The fluctuations on $N$, $S_p$ and $\theta_p$ are described by adding Langevin noise sources $F_N(t)$, $F_{Sp}(t)$ and $F_{\theta p}(t)$ to rate equations (5) – (7), respectively. These noise sources have Gaussian probability distributions with zero mean values. Also, they are $\delta$-correlated processes with the correlations [5,16]

$$\langle F_a(t) F_b(t') \rangle = V_{ab} \delta(t-t') \tag{11}$$

where $a$ and $b$ stand for any of $N$, $S_p$ or $\theta_p$ and the correlation variances $V_{ab}$ are determined from the steady state solutions of the rate equations at each integration step. It should be noted that the orthogonally among the oscillating modes as well as between the mode intensity and phase require $V_{SpSq} = V_{SpSq}\delta_{pq}$ and $V_{Sp\theta q}=0$.

The frequency content of the fluctuating mode intensity is measured in terms of RIN using the time fluctuations $\delta S(t) = S(t) - \overline{S}$ in the total photon number $S(t) = \sum_p S_p(t)$ around its average value $\overline{S}$. Over a finite time $T$, RIN is calculated from the Fourier transformation [16]

$$RIN = \frac{1}{\overline{S}^2}\left\{\frac{1}{T}\left|\int_0^T \delta S(t) e^{-j2\pi f\tau} d\tau\right|^2\right\} \qquad (12)$$

where $f$ is the Fourier frequency. Similarly, RIN of the individual modes is defined in terms of the fluctuations of their photon number $S_p(t)$.

Definitions of the other laser parameters in the above equations and their typical values in InGaAsP laser are given in Table 1.

**Table 1.** Definition and typical values of the parameters of the FP-InGaAsP laser used in present calculations.

| Parameter | Meaning | Value | Unit |
|---|---|---|---|
| $\lambda_{peak}$ | Emission wavelength | 1.55 | Mm |
| $a$ | Deferential gain coefficient | $8.24 \times 10^{-12}$ | $m^3 s^{-1}$ |
| $\xi$ | Field confinement factor in the active layer | 0.15 | -- |
| $V$ | Volume of the active region | $30 \times 10^{-18}$ | $m^3$ |
| $N_g$ | Electron number at transparency | $3.69 \times 10^7$ | -- |
| $b$ | Dispersion parameter of the linear gain spectrum | $5.07 \times 10^{37}$ | $m^2 A^{-2}$ |
| $B$ | Coefficient of self-suppression of modal gain | $9.02 \times 10^{-7}$ | $s^{-1}$ |
| $\tau_{in}$ | Electron intraband relaxation time | 0.1 | ps |
| $N_s$ | Electron number characterizing gain suppression | $1.01 \times 10^8$ | -- |
| $\alpha$ | Linewidth enhancement factor | 3.5 | -- |
| $\tau_s$ | Electron lifetime by spontaneous | 0.779 | ns |
| $\Delta\lambda$ | Half-width of spontaneous emission | 23 | nm |
| $\tau_p$ | Photon lifetime | 1.69 | ps |
| $I_{th}$ | Threshold current | 25 | mA |

### 3. Methodology of Numerical Calculations

We count 15 modes ($p = -7 \rightarrow +7$) in the calculations and integrate the 16- rate equations (5) – (7) numerically using the fourth order Runge-Kutta method. The time step of integration is as short as $\Delta t = 0.1$ ps. The parameters of 1550-nm InGaAsP laser are listed in table 1. This laser is assumed coupled with a very-short external cavity of length $L_{ex} = 2.5$ mm, which corresponds to round trip propagation delay time of $\tau = 16.7$ ps in the feedback cavity. These parameters in the single-mode version were predicted to support modulation bandwidth as wide as 55 GHz [11,17]. SMSR of the multimode oscillation is evaluated as the ratio of the average value of the power of the dominant mode to that of the strongest side mode. We decide the laser oscillation to be in the single mode operation when SMSR = 100 or 20 dB.

At each integration instant $t_i$, the noise sources $F_{Sp}(t_i)$ and $F_N(t_i)$ are generated using the following forms [8].

$$F_{Sp}(t_i) = \sqrt{\frac{V_{SpSp}}{\Delta t}} g_{Sp} \tag{13}$$

$$F_{\theta p}(t_i) = \frac{1}{2[S(t_{i-1})+1]} \sqrt{\frac{V_{SpSp}}{\Delta t}} g_{\theta p} \tag{14}$$

$$F_N(t_i) = \sqrt{\frac{V_{NN}(t_i) + 2\sum_p k_p(t_i) V_{SpN}(t_i)}{\Delta t}} g_N + \sum_p \frac{V_{SpSp}(t_i)}{V_{NSp}(t_i)} \{F_{Sp}(t_{i-1}) + 2[S(t_{i-1})+1] F_{\theta p}(t_{i-1})\} \tag{15}$$

The variances $V_{ab}$ (with $a$ and $b$ referring to each of the symbols N, $S_p$ or $\theta_p$) at time $t_i$ are evaluated from $S(t_i-1)$ and $N(t_i-1)$ at the preceding time $t_i-1$ by assuming quasi-steady states ($dS_p/dt \approx dN/dt \approx 0$) over the integration step $\Delta t = t_{i-1} - t_i$,

$$V_{SpSp}(t_i) = 2\left[\frac{a\xi}{V} S_p(t_{i-1}) + \frac{C_p}{\tau_s}\right] N(t_{i-1}) \tag{16}$$

$$V_{NN}(t_i) = 2\left[\frac{1}{\tau_s} + \frac{a\xi}{V} \sum_p S_p(t_{i-1})\right] N(t_{i-1}) \tag{17}$$

$$V_{NSp}(t_i) = -\frac{a\xi}{V}\left[N(t_{i-1}) - N_g\right] S(t_{i-1}) - \frac{N(t_{i-1})}{\tau_s} \tag{18}$$

In Eqs. (14) – (16), $g_N$, $g_{sp}$ and $g_{\theta p}$ are independent Gaussian random numbers with variances of unity and zero mean values and are obtained by applying the Box–Mueller approximation [18] to a set of uniformly distributed random numbers generated by the computer.

The calculation of RIN via equation (12) is computed from the obtained values of $S(t_i)$ at the $i^{th}$ integration step by employing the fast Fourier transform (FFT) as

$$RIN = \frac{1}{\bar{S}^2} \frac{\Delta t^2}{T} |FFT[\delta S(t_i)]|^2 \tag{19}$$

### 4. Results and Discussions

At beginning we gain inside into influence of both AGS and intensity fluctuations on the dynamics of the multimode laser. Figure 1(a) plots the time trajectories of the strongest three modes $p = +4, +2$ and 0 as well as of the total output at times much longer than the transient region. In this case the noise sources are dropped from rate equations (5) – (7). The figure shows that the modes reach the steady state and the laser output is mainly contained in the three modes. In this case SMSR = $S_{+1}/S_{+2}$ = 1.1, supporting the multimode oscillation character of the laser. It is worth noting that one of these modes is the central one and the other two modes lie on the longer-wavelength side because of AGS that works to increase the gain of longer-wavelength modes while suppressing the gain of modes with shorter wavelengths [14]. AGS induces strong coupling and competition among these long-wavelength modes [5].

Figure 1(b) plots the corresponding time trajectories of $S_p(t)$ and total output $S(t)$ as solutions of the stochastic rate equations (5) – (7). The figure shows the typical hopping multimode oscillations in the output of long-wavelength laser in the time domain [7]. The oscillating modes do not reach the steady state, instead they exhibit switching to the lasing level in a semi-periodic manner, $p = +2 \rightarrow +3 \rightarrow +4 \rightarrow 1$, etc. That is, the mode coupling is in the form of anti-correlations among these hopping modes [6]. The frequency of this mode hopping or switching is $f_{MH} = 420$ MHz. The interpretation of this multimode hopping is that AGS induces a bistable state in the mode dynamics, and the noise sources work as triggers to the oscillating modes to switch between this bi-stable state [5].

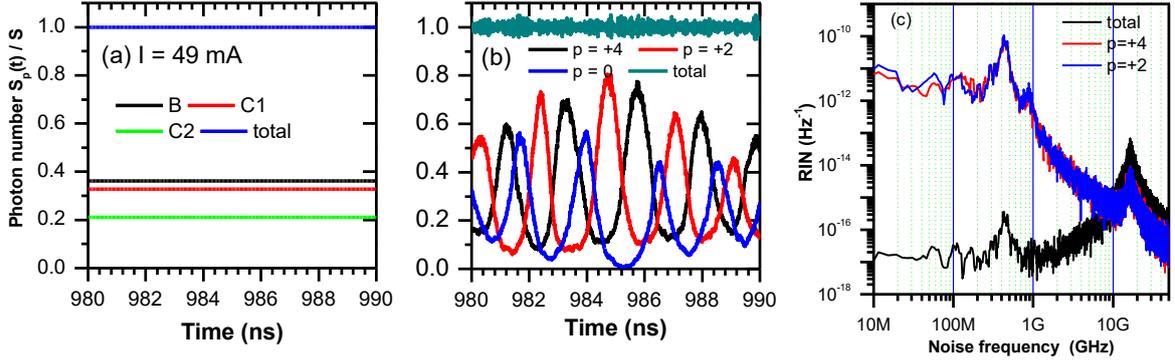

**Figure 1.** Time-domain characteristics of the solitary multimode laser (a) without and (b) with noise triggering when $I = 49$ mA. (c) The laser attains steady state operation without noise whereas the noise induced multimode hopping.

The intrinsic intensity fluctuations is seen in the figure as fluctuations of the switching state of the hopping modes as well as fluctuations of the total intensity around the steady-state value. The corresponding frequency spectrum of RIN of the total output as well as of the strongest two modes $p = +4$ and $+2$ are given in figure 1(c). The spectra exhibit a sharp peak around the carrier-photon resonance (CPR) frequency of the laser, $f_r = 15$ GHz. The low-frequency part of RIN of the oscillating modes is much higher than that of the total RIN, a feature known as mode-partition noise [1]. This difference of noise is almost six orders of magnitude. The low-frequency RIN of the total output is LF-RIN = $2 \times 10^{-16}$ Hz$^{-1}$, while it is $1 \times 10^{-11}$ Hz$^{-1}$ for the hopping modes. Also, the figure displays the interesting multimode hopping peak**s** around 420 MHz, which corresponds to the frequency of semi-periodic switching of the hopping modes $f_{MH}$ to the lasing level seen in figure 1(a). Such peaks were observed in the measured RIN spectrum of InGaAsP lasers [8].

Influence of OFB from the very close reflector on the modal oscillation of the investigated laser is depicted in figure 2. The figure plots SMSR versus the external reflectivity $\Gamma R_{ex}$ (measure of the coupled light to the primary laser cavity) normalized by the front facet reflectivity. It is worth noting that the laser operates in CW over the relevant range of $\Gamma R_{ex}/R_f$, which is a typical advantage of using the very short external cavity in avoiding the noisiest state of chaos [10]. The figure shows that over the relevant range of the OFB strength, the multimode oscillation of the laser is converted into oscillation in single mode over five regimes of optical feedback: namely, $0.0075 \leq \Gamma R_{ex}/R_f \leq 0.025$, $0.039 \leq \Gamma R_{ex}/R_f \leq 0.083$, $0.114 \leq \Gamma R_{ex}/R_f \leq 0.188$, $0.295 \leq \Gamma R_{ex}/R_f \leq 0.363$ and $0.467 \leq \Gamma R_{ex}/R_f \leq 1.01$. The characteristics of the RIN spectra and modal oscillation in the investigated regimes of single and multimode are illustrated in the following series of figures.

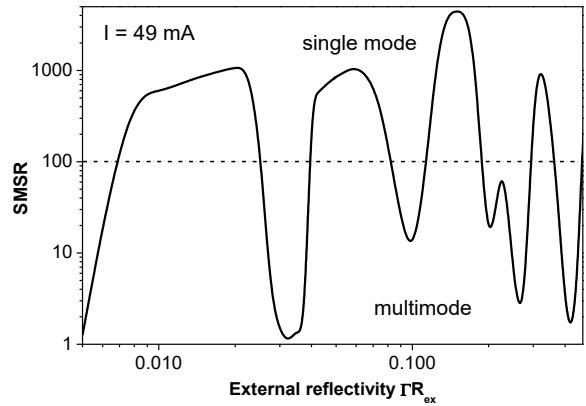

**Figure 2.** Variation of SMSR with external reflectivity $\Gamma R_{ex}$ when $I = 49$ mA. Regimes of single and multimode oscillations under OFB are specified.

In figures 3(a-c), we characterize the multimode oscillations at three strengths of feedback of $\Gamma R_{ex}/R_f = 0.035$, 0.25, and 0.45 by plotting the corresponding output spectra, which confirm the multimode oscillation character of the laser indicating values of SMSR = 1.45, 4.7 and 6.3 oscillation, respectively. In figure 3(a), the strongest three modes are $p = 0, +1$ and $+4$, while in figure 3(b), these modes are $p = -2, -1$ and $+3$ and in figure 3(c) these modes become $p = -5, 0, +3$. That is, modes with shorter wavelengths attain stronger power and share the total laser output with the increase of OFB. This result agrees with the prediction of Ahmed and Yamada [19, 20] that OFB may act against AGS and increase the gain of shorter wavelength side modes so as to increase their share in the laser output.

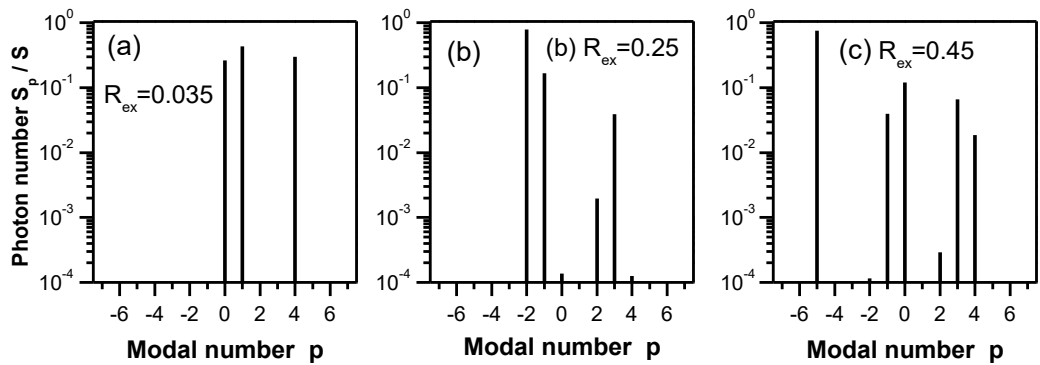

**Figure 3.** Examples of output multimode spectra of the laser under OFB of $\Gamma R_{ex} = 0.035$, 0.25 and 0.45. The strongest modes shift to shorter wavelengths with the increase of feedback.

In figures 4(a-c), we plot the spectral characteristics of RIN of the total output and the strongest two modes that correspond to the hopping multimode oscillations in figures 3(a-c), respectively. The RIN spectra exhibit the CPR peaks around the CPR frequency $f_r = 14$GHz of the laser. The figure shows also that these RIN spectra reveal another peak around the mode-hopping frequency in the low-frequency regime. This frequency is $f_{MH} = 330$MHz when $\Gamma R_{ex} = 0.035$, which decreases to 50MHz when $\Gamma R_{ex} = 0.25$ and to 102MHz when $\Gamma R_{ex}$ increases to 0.45. In figures 4(a) and (b), the noise spectrum is flat and superimposed by the mode-hopping peak. The flatness of the RIN spectrum was attributed to the phase distortion between the internal reflected light and the external feed-backed light [12,21]. Cause of this white noise is also explained in terms of competition among external modes whose lasing frequency is decided by the space between the laser facet and the reflecting mirror [21 - 23]. In figure 4(a) of $\Gamma R_{ex} = 0.035$, the total RIN is almost flat except for the MC peak reveling RIN = $8.2 \times 10^{-16}$ Hz$^{-1}$ at $f = 100$MHz while it is $2 \times 10^{-12}$ and $1.8 \times 10^{-12}$ Hz$^{-1}$ for the hopping modes $p = +1$ and $+4$. In figure 4(b) when $\Gamma R_{ex} = 0.25$, RIN is higher for both the total RIN and modal RIN with levels $7.9 \times 10^{-15}$Hz$^{-1}$ and $1.1 \times 10^{-11}$ Hz$^{-1}$, respectively. The corresponding high-frequency peak occurs at higher frequencies of 37 and 40 GHz frequency as a result of increase of OFB strength. This peak corresponds to high-frequency oscillations which is a trigger of high-frequency photon-photon resonance (PPR) when this laser is modulated with sinusoidal modulation with comparable frequencies. On the other hand, the RIN spectrum in figure 4(c) is of the low-frequency type with $1/f$ dependence as reported by Imran and Yamada [13] who attributed thus noise to the strong mode competition induced by OFB among the lasing modes in the solitary laser. In the low-frequency noise, the total RIN and modal RIN are much enhanced to the orders of $10^{-10} \sim 10^{-8}$ Hz$^{-1}$. These noise characteristics are in good correspondence with those predicted in theory and measured in experiments [10, 12, 15, 20, 23, 24].

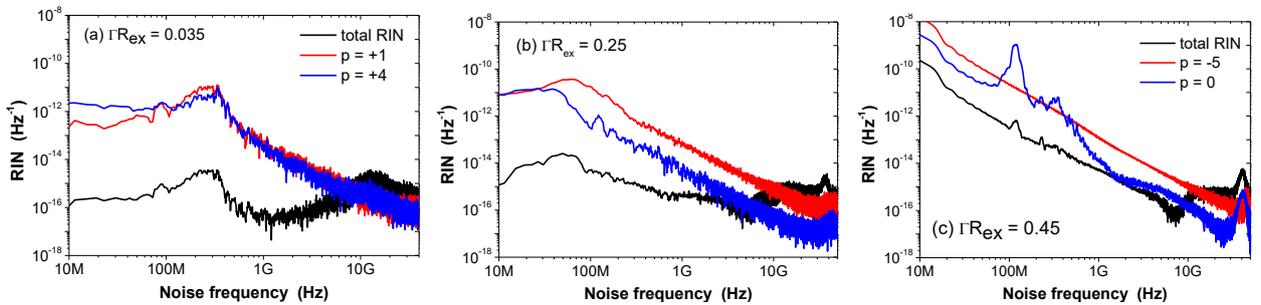

**Figure 4.** RIN spectra of the total output and strongest two modes characterizing the output multimode spectra in figure 3. The mode hopping frequency increases with the increase of OFB and the low-frequency levels is more enhanced.

In figure 5(a-c), we characterize the output spectra under SMO which are achieved at the three strengths of feedback of $\Gamma R_{ex}/R_f$ = 0.02, 0.15, and 0.325. These figures confirm that the laser output is mainly contained in one mode with SMSR = 1080, 4807 and 1294, respectively. Also, the figures indicate jumping of the predominant mode in the long-wavelength side from $p = 0$ to $p = +2$ and to $p = +3$, with the increase of the OFB strength. Since this effect is manifestation of the AGS [8], OFB works in this case to sustain AGS such that one of long-wavelength modes achieves highest gain and predominates the laser oscillations.

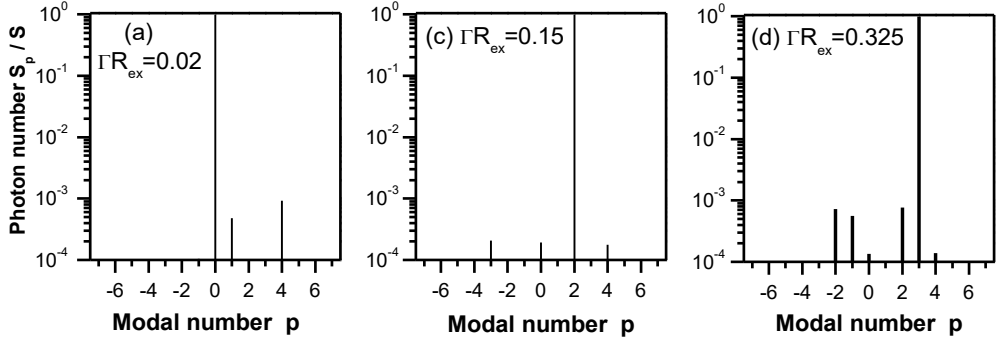

**Figure 5.** Examples of output single-mode spectra of the laser under OFB of $\Gamma R_{ex}$ = 0.02, 0.15 and 0.325. The dominant mode shifts to longer wavelengths with the increase of feedback.

Figures 6 (a-c) characterize the RIN spectra of SMO in figures 5 (a-c), respectively. The figures indicate flat (white) noise in the low-frequency regime and disappearance of the mode-hopping peak. That is, the modes attain a steady state in the time domain and the laser does not exhibit the multimode hopping. This result confirms the conclusion of figure 5 that AGS and OFB work to enhance the gain of one-long-wavelength mode over the other modes to predominate the laser oscillations. The RIN level when $f$ = 100MHz is $1.3 \times 10^{-17}$ $Hz^{-1}$ when $\Gamma R_{ex}$=0.02, which increases to $8.8 \times 10^{-16}$ $Hz^{-1}$ and $4.0 \times 10^{-15}$ $Hz^{-1}$ when $\Gamma R_{ex}$ increases to 0.15 and 0.325, respectively. Therefore, the increase of the OFB strength results in stronger phase distortion between the internal field and the OFB field in the laser cavity with the increase [21-23]. The figures display also the high-frequency peak of the RIN spectrum. In figure 6(a) of weak OFB, this peak still corresponds to the CPR of the laser around $f$ = 15 GHz, while in figures 6(b) and (c) of stronger OFB, this peak corresponds to PPR and periodic oscillation of one of the external-cavity modes induced by OFB [11]. In figure 6(c), this oscillation frequency is as high as 40 GHz, which indicates possibility of achieving high-frequency/speed modulation with single-mode oscillation of the laser under strong OFB due to PPR [25].

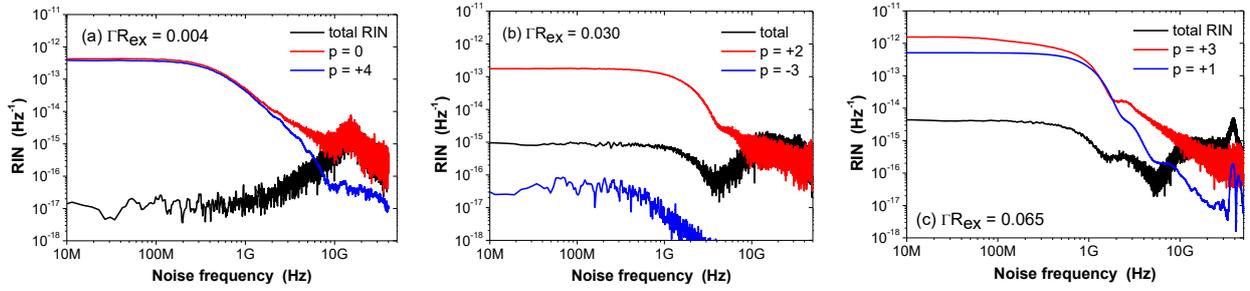

**Figure 6.** RIN spectra of the total output and strongest two modes characterizing the output SMO spectra in figure 5. The low-frequency levels are flat and increase with the increase of OFB.

Influence of OFB on LF-RIN of the total laser output and of the strongest oscillating mode is investigated in figure 7 over the relevant range of $\Gamma R_{ex}$. The dashed regimes correspond to single-mode oscillations. The modal noise is higher than the total RIN, indicating the phenomenon of mode partition. The figure indicates that LF-RIN is enhanced in the regimes of multimode oscillations due to the mode hopping. In the regimes of single-mode oscillations, RIN decreases with the increases of OFB; that is OFB works to improve the noise performance of the laser when it achieves SMO.

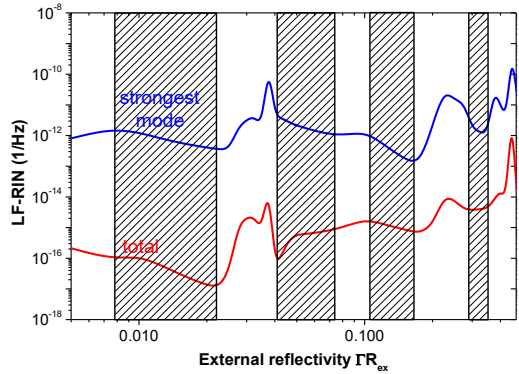

**Figure 7.** Variation of LF-RIN with external reflectivity $\Gamma R_{ex}$ when $I = 49$ mA. The RIN level increase with the increase of OFB. The multimode hopping makes the noise level in the regimes of multimode much higher than that of the single-mode oscillation.

Prior to concluding, we will take the moment to discuss some insights into brining the work of this paper into a larger context. Here is interesting to ask what possibilities arise when considering single mode (or few-mode) cavities [27-29] and laser physics [30-37], both at the nanoscale and (sub) diffraction limited optical modes. Interestingly, the debate around whether, or not, the Purcell factor, which captures the light-matter-interaction strength such as of a laser cavity and is proportional to the cold-cavities' quality factor (Q) divided by the cavities' mode volume, has an influence on both the ASG and the gain relaxation frequencies, i.e. speed of the laser [30]. Indeed, the Purcell factor is especially high in light emitters and lasers featuring a sub diffraction limited optical mode, primarily due to the nonlinear scaling of volume and introduced loss due to the cavities' inability to provide feedback. This impacts the ASG in such small-volume cavities at (or below) the diffraction-limit of light [30-33], because the laser design with enhanced $F_p$ are also capable of increasing the temporal RO of the laser cavity, thus expanding the 'speed' of the laser under direct modulation. For instance, an example of this is the transverse-coupled cavity laser design providing coherent feedback from a plurality of cavities, thus enhancing the light emission from a central lasing cavity [38, 40-45]. Looking ahead, future research should also investigate the effects of ASG in cavities with high longitudinal modes, for example in fiber optic-based laser systems for Brillouin amplification [39].

**Conclusions**

We showed that OFB from a very short external cavity can convert HMMO that characterizes long-wavelength semiconductor laser into SMO and improve the noise performance of the laser. The study was based on a modified time-delay multimode rate-equation model that includes modal gain suppression mechanism. The modal oscillation of the laser depends on the manner how OFB affects the cross modal AGS. OFB may increase AGS to enhance the gain of one longer wavelength mode and then convert the multimode oscillation into predominant oscillations in that mode. When OFB works to release AGS, it sustains hopping oscillation among long–wavelength modes in additions to other modes on the short-wavelength side of the gain spectrum. The total RIN and mode-partition noise are enhanced in the regimes of HMO, whereas RIN decreases with the increases of OFB in the regimes of SMO.

**Acknowledgment**

This project was funded by the Deanship of Scientific Research (DSR) at King Abdulaziz University, Jeddah, under grant no. (**RG-18-130-41**). The authors, therefore, acknowledge with thanks DSR technical and financial support.